\documentstyle[prl,aps,epsf,twocolumn]{revtex}
\begin{document}
\draft
\twocolumn[\hsize\textwidth\columnwidth\hsize\csname@twocolumnfalse%
\endcsname

\preprint{}
\title{Quasiparticle transport and localization in high-$T_c$ superconductors}
\author{T. Senthil, Matthew P.A. Fisher, Leon Balents,\cite{lucent} 
Chetan Nayak\cite{ucla}}
\address{ Institute for Theoretical Physics, 
University of California,
Santa Barbara, CA 93106--4030  
}
 
\date{\today}
\maketitle
 
\begin{abstract}
We present a theory of the effects of impurity scattering in
$d_{x^2-y^2}$ superconductors and their quantum disordered
counterparts, based on a non-linear sigma model formulation.  We show
the existence, in a quasi-two-dimensional system, of a novel
spin--metal phase with a non-zero spin diffusion constant at zero
temperature.  With decreasing inter-layer coupling, the system
undergoes a quantum phase transition (in a new universality
class) to a localized spin-insulator.  Experimental implications for
spin and thermal transport in the high-temperature superconductors are
discussed.
\end{abstract}
\vspace{0.15cm}

%\pacs{PACS numbers:75.10.Nr, 05.50.+q, 75.10.Jm}
%\vskip -0.5 truein
]
\narrowtext
%\newpage

Over the last few years, experiments\cite{Kirtley} have convincingly
established that the superconducting state of the hole-doped cuprate
materials is characterized by spin singlet $d_{x^2-y^2}$ pairing. In
such a superconductor, the gap vanishes at four points on the
(two-dimensional) Fermi surface. The quasiparticle excitations at
these ``nodes'' have a linear dispersion, and an associated density of
states that vanishes linearly on approaching the Fermi surface. This
leads to power law dependences of various physical quantities on
temperature.
Impurity scattering is
expected to strongly modify these properties. Experimentally, the
power law temperature dependences are rounded off, apparently
approaching constant, temperature-independent behavior at the lowest
temperatures. This fact is well reproduced by approximate,
self-consistent treatments of impurity scattering which show that a
constant finite density of states is generated at the Fermi energy for
any arbitrarily weak impurity strength
\cite{Lee,Gorkov}. Quasiparticle transport properties 
have also been investigated\cite{HPS,Graf}
theoretically with such self-consistent approximations with some
phenomenological success. Going further, Lee\cite{Lee} has suggested,
on the basis of calculations of the zero frequency microwave
conductivity, that the quasiparticle eigenstates are strongly
localized.

In this paper, we reconsider the effects of disorder on the low energy
quasiparticles in the $d_{x^2-y^2}$ superconductor. We point out that
the problem of quasiparticle transport and localization in a
superconductor is conceptually very different from the more familiar
situation of non-interacting electrons in a random potential. This is
because, unlike in a normal metal, the charge of the quasiparticles in
the superconductor is not a conserved quantity. This immediately
implies that the quasiparticle charge in the superconductor cannot be
transported through diffusion. Indeed the quasiparticle charge density
is {\em not} a hydrodynamic mode in the superconductor. However, in a
singlet superconductor (and in particular in the high-$T_c$
superconductors), the condensate does not carry any spin, and
consequently the spin of the quasiparticles is a good quantum number
and is conserved.  The quasiparticle energy is also
conserved. Thus, there is the possibility of having spin and energy
diffusion without charge diffusion. These differences in symmetry lead
to interesting differences between the localization properties of
quasiparticles in the superconductor, and in the normal metal.
Such differences have been pointed out before\cite{AZ} in the 
context of the random matrix theory of mesoscopic
normal/superconducting systems.

We address quasiparticle transport using a replica field theoretic
formulation. As expected, the field theory is different from that
describing Anderson localization in a normal metal.  The properties of
the theory are determined by a single coupling constant, which is the
dimensionless {\em spin} conductance. This is the physically correct
quantity whose behavior as a function of system size enables
construction of a scaling theory of localization.  By analyzing the
properties of the field theory, we show the existence of a logarithmic
``weak localization" correction in two dimensions suggesting
localization at the largest length scales.  This correction persists, in part,
in the presence of an orbital magnetic field (unlike usual Anderson
localization) or a Zeeman field, but is suppressed when both are
present.  In all cases, however, the quasiparticles are generically
ultimately localized in two dimensions.  Upon inclusion of interlayer
coupling, there is the interesting possibility of a quantum phase
transition between an extended {\sl spin metal} and a
localized  {\sl spin insulator}.  The 
spin metal has diffusive spin correlations, a
finite spin susceptibility, and an associated finite spin conductivity
all {\sl at zero temperature}.  We are not aware of the existence of
such a spin phase in any insulating Heisenberg spin
model with or without randomness. 
  
The spin insulator is expected to exhibit local moments and spin-glass
or random-singlet behavior at very low temperatures.  The transition
between these two phases is described by the critical point of the
replica field theory (neglecting quasiparticle interactions), and is a
new universality class for localization.

Most of these results also go over unmodified to the quantum
disordered version of the $d_{x^2-y^2}$ superconductor - the ``nodal
liquid" phase that has been analyzed recently\cite{BFN} as a possible
low temperature theory of the pseudo-gap regime in the cuprate
materials.  

We begin our analysis with the lattice quasiparticle Hamiltonian
for a singlet superconductor,
\begin{equation}
\label{BCS_c}
    {\cal H} =
    \sum_{i,j}\left[t_{ij}\sum_{\sigma}c^{\dagger}_{i\sigma}c_{j\sigma} + 
    \Delta_{ij}c^{\dagger}_{i\uparrow}c^{\dagger}_{j\downarrow}+ 
    \Delta^{*}_{ij}c_{j\downarrow}c_{i\uparrow} \right]
\end{equation}
where $i,j$ are site labels.  Using Hermiticity combined with
spin-rotational and time-reversal invariances, $t$ and $\Delta$ may be
taken to be real-symmetric matrices.  Note that the total number of
particles is not conserved by this Hamiltonian while their total spin
is.  In the presence of impurities, we define $t_{ij} =
t^{0}_{ij} + t^{1}_{ij}$ and $\Delta_{ij} = \Delta^{0}_{ij} +
\Delta^{1}_{ij}$, where $t^0$ and $\Delta^0$ are the Fourier
transforms of the kinetic energy $\epsilon_k$ (measured from the Fermi
energy) and the gap function $\Delta_k$, respectively.  In the
$d_{x^2-y^2}$ superconductor of primary interest, we may take
$\Delta_k = \Delta_0 (\cos k_x - \cos k_y)$.  We mention in passing
that for weak impurities, a continuum limit may be taken, focusing on
wavevectors near the d-wave nodes.  The resulting ``dirty Dirac''
Hamiltonian is similar to various models in the
literature\cite{Tsvelik}, but differs from previously studied variants
in that it contains several random {\sl anomalous} couplings. The 
Hamiltonian above can be regarded as a lattice regularization of 
this continuum effective field theory for the $d_{x^2-y^2}$
superconductor, hence our results are quite general and not 
restricted to a BCS approximation.

The effect of weak
randomness can be analyzed by perturbative renormalization group
calculations\cite{long}\ which show that the randomness is a
(marginally) relevant perturbation.  To make progress then, we employ
a field-theoretic reformulation of the self-consistent treatment
adopted in earlier works on dirty d-wave superconductivity\cite{Lee}.
This begins with the standard coherent-state functional integral
formulation, in which the electron operators $c,c^\dagger$ are
replaced by Grassman fields $c,\overline{c}$ averaged with respect to
a statistical weight $e^{-S}$, where $S$ is an action.  As the
randomness is independent of time and ${\cal H}$ is quadratic,
different pairs of frequencies $(\omega,-\omega)$ decouple, and it is
sufficient for our purposes to focus simply on $\omega=0$.

Several notational conventions are convenient.  We define
four-component fields $\psi_{ia\alpha}$, with $\psi_{i1\alpha}
\equiv c_{i\alpha}/\sqrt{2}$ and $\psi_{i2\alpha} \equiv
i\sigma^y_{\alpha\beta}\overline{c}_{i\beta}/\sqrt{2}$.  From this point on
we adopt a notation in which $\vec{\tau}$ and $\vec{\sigma}$ matrices
act in the particle/hole ($a$) and spin ($\alpha$) spaces,
respectively.  A conjugate field is then defined by
$\overline{\psi}_i = (C\psi_i)^T$, where $C=\sigma^y\tau^y$.
The action in these variables appears non-anomalous,
\begin{equation}
    S = \sum_{ij} \overline{\psi}_i \left( t_{ij}\tau^z +
    \Delta_{ij}\tau^x\right)\psi_j + i\eta \sum_i
    \overline{\psi}_i\sigma^z\psi_i.
    \label{Spsi}
\end{equation}
At this stage we have also included an infinitesimal imaginary Zeeman
field $\eta$, which acts to generate physical correlation functions.

To compute disorder-averaged quantities, we replicate the fields $\psi
\rightarrow \psi^\mu$, with $\mu=1\ldots n$, so that for $n
\rightarrow 0$ the statistical weight is normalized for each
realization of the randomness.  Physical quantities can now be simply
expressed.  In particular, the spin susceptibility is $\chi_0 =
-(1/\pi) {\rm Im}\langle
\overline{\psi}_i^\mu\sigma^z\psi_i^\mu\rangle$ (no sums).  Angular
brackets denote both field-theoretic ($\psi$) and disorder averages.
The spin diffusion constant, $D_s$, can also be determined from the
``diffusion propagator'' $P_{ij}$, whose Fourier transform is $P(q) =
\sum_j P_{ij}\exp[\vec{q}\cdot(\vec{x}_i-\vec{x}_j)] = 8\pi\chi_0/(D_s
q^2)$.  One has $P_{ij} = -\langle (\overline{\psi}^\mu_i \sigma^+
\psi_i^\nu) (\overline{\psi}^\nu_j\sigma^-\psi_j^\mu)\rangle$, with
$\sigma^\pm = (\sigma^x \pm i \sigma^y)/2$ and no replica sum should
be taken.

The ensemble average over $t^1,\Delta^1$ can now be immediately
performed, generating a translationally-invariant action with
non-trivial quartic couplings between different replicas.  A more
general analysis\cite{long}\ demonstrates that the essential features
are captured by uncorrelated zero-mean local Gaussian fields $t^1_{ij}
= t_i^1\delta_{ij}$ and $\Delta^1_{ij} = \Delta_i^1\delta_{ij}$ with
covariances $[t_i^1t_j^1]_{\rm ens.} = [\Delta^1_i\Delta^1_j]_{\rm
ens.} = u\delta_{ij}$.  With this choice, the algebra is particularly
simple, and the quartic interactions can in turn be decoupled via two
$2n\times 2n$ Hermitian Hubbard-Stratonovich fields, $Q$ and $P$,
acting in the spin and replica spaces (diagonal in the particle/hole
space).  The effective action becomes
\begin{eqnarray}
    {\cal S} & = & \sum_i \frac{1}{u} {\rm Tr} \left[(Q(i))^2 +(P(i)
    )^2\right] +
    2\sum_{ij}\overline{\psi}_i \Big[ \big(iQ(i) 
    \nonumber \\ & & 
    - P(i)\tau^y +i\eta\sigma^z\big)\delta_{ij} + t^0_{ij}\tau^z +
    \Delta^0_{ij}\tau^x\Big]\psi_j,
    \label{SPQ}
\end{eqnarray}
where we have suppressed spin and replica indices.

A saddle-point (in $Q$ and $P$) analysis of Eq.~\ref{SPQ}\ recovers
the conventional self-consistent approximation.  In particular, one
finds $Q = 2\pi \chi_0 \sigma^z$ and $P=0$.  The constant
$\chi_0$ appears as an imaginary self-energy, is (the saddle-point
approximation to) the physical spin-susceptibility, and represents a
generation of a non-zero ``quasiparticle density of states (DOS)'' due
to disorder.  The conclusion that $\chi_0 \neq 0$ is amply supported
by experiment, leading us to believe that this saddle-point is a
physically correct starting point.  The imaginary self-energy also has
a complementary interpretation as a finite (inverse) elastic
scattering time $1/\tau_e$.  For times longer than $\tau_e$,
quasiparticles no longer move ballistically, and we expect diffusion
{\sl of the conserved energy and spin densities}.

Fluctuations around this saddle-point represent both diffusion and
corrections to it.  Near two spatial dimensions these fluctuations are
captured by a Non-Linear Sigma-Model (NL$\sigma$M) treatment.  The crucial
ingredients are the physical (non-statistical) symmetry properties of
the Hamiltonian, which determine symmetries of the replicated
action, Eqs.~\ref{SPQ}.  For the SU(2) and time-reversal invariant
form chosen, the crucial symmetry group is $Sp(2n)\times Sp(2n)$.  In
particular, consider the transformation $\psi_i
\rightarrow U \psi_i$, with $U={1
\over 2}[U_A(1+\tau^y) + U_B(1-\tau^y)]$, with $U_{A,B}$ $2n\times 2n$
unitary matrices in the spin and replica spaces satisfying
$U_{A,B}^T\sigma^y U_{A,B} = \sigma^y$.  Under this transformation,
the other fields rotate according to $Q+iP \rightarrow U_A^\dagger
(Q+iP) U_B$.  For $\eta=0$, all such rotations leave ${\cal S}$
invariant, while this is true for non-zero $\eta$ only when
$U_B=\sigma^z U_A \sigma^z$, hence $\eta$ breaks the symmetry
infinitesimally from $Sp(2n)\times Sp(2n)$ to $Sp(2n)$.  For a single
replica, note that as
$Sp(2) \simeq SU(2)$, one of these $Sp(2)$ symmetries
is just spin rotation invariance. The other $Sp(2)$ symmetry is
actually a consequence of time reversal invariance, and can be traced
to the reality of the Hamiltonian $H$. 

The NL$\sigma$M is constructed by considering fluctuating
$Sp(2n)\times Sp(2n)$ rotations of saddle-point solutions that are
slowly-varying in space.  In general, these can be shown to take the form
of an $Sp(2n)$ matrix $U(\vec{x})$, with $Q(x) + i P(x) = {\pi \over
2}\chi_0 \sigma^zU(\vec{x})$.  The form of the action for $U$ is determined
entirely on symmetry grounds, and is verified by a direct
calculation\cite{long}\ expanding $Q$ and $P$ and integrating out
non-critical massive modes.  We find
\begin{equation}
\label{chiral}
S_{{\rm NL}\sigma{\rm M}} = \int d^2x \frac{1}{2g}{\rm Tr}\,(\nabla U \cdot
\nabla U^{\dagger}) -\eta {\rm Tr}\,(U + U^{\dagger})
\end{equation}
where $U(x) \in Sp(2n)$.  This field theory is known as the
``principal chiral $Sp(2n)$ model'' in the field theory literature. In
contrast to the conventional sigma models used to describe the
localization of non-interacting electrons, here the field variables
live on a group manifold instead of a coset space.  The $Sp(2n)\times
Sp(2n)$ symmetry acts on $U$ via global left and right multiplication
with independent $Sp(2n)$ matrices.

The replica-diagonal self-consistent approximation used in other work
corresponds to keeping only the configuration $U(x) = \bbox{1}$ in the
action. Small quadratic fluctuations around this solution correspond
to diffusion, and a direct calculation of $P_{ij}$ in this
approximation\cite{long}\ relates the coupling-constant to the
spin-conductance $\sigma_s$, to wit $\frac{1}{g} = \frac{\pi}{2}
\sigma_s$.  The derivation of the sigma model provides an estimate for
the bare coupling constant: $\frac{1}{g_0} = \frac{1}{4\pi}\frac{v_F^2
+ v_{\scriptscriptstyle\Delta}^2}{v_F v_{\scriptscriptstyle\Delta}}$
with $v_F$, the Fermi velocity, and $v_{\scriptscriptstyle\Delta}$,
the slope of the $d_{x^2-y^2}$ gap linearized near the nodes.  Note
that this is independent of the disorder strength. A similar result
for the zero frequency microwave conductance was obtained earlier by
Lee\cite{Lee}, in particular $\sigma(\omega=0^+ ) = {1 \over
\pi^2}(v_F/v_{\scriptscriptstyle\Delta}) e^2/h$.  The difference in
the velocity-dependence of the prefactors is conceptually significant:
the spin-conductance obeys an Einstein relation while the microwave
conductance cannot.  This distinction arises because the quasiparticle
charge is not a good quantum number.

Consider separately the orbital and Zeeman couplings to an applied
magnetic field.  The orbital field breaks time-reversal symmetry but
not $SU(2)$, and similar manipulations to those above lead ultimately
to a $Sp(2n)/U(n)$ NL$\sigma$M, also distinct from the three
conventional universality classes of dirty metals.  The Zeeman
coupling, by contrast, breaks $SU(2)$ invariance, leaving only a
$U(1)$ spin-rotation symmetry around the field axis (say $\hat{\bf
z}$).  Using the particle/hole transformation,
$c^{\vphantom\dagger}_\downarrow \rightarrow c^\dagger_\downarrow$,
this $U(1)$ symmetry is easily shown to play the same role as does the
charge-conservation $U(1)$ in conventional localization.
Consequently, the Zeeman field leads to the usual orthogonal sigma
model, and Zeeman and orbital effects together drive the system to the
unitary universality class.

All these field theories exhibit diffusion on length scales of order
the elastic mean free path $\ell_e$.  Beyond this scale, quantum
interference corrections can play an important role.  They
are determined by the renormalization group (RG) equation for $g$, which
for the $Sp(2n)$ and $Sp(2n)/U(n)$ sigma models can be found in
Refs.~\onlinecite{Polyakov}:
\begin{equation}
    \frac{dg}{d\ln L} = -\epsilon g +\frac{\alpha g^2}{4\pi} + O(g^3)
    \label{rgeqn}
\end{equation}
where $\epsilon = d -2$ and we have set $n = 0$. The number $\alpha = 1,
\frac{1}{2}$ for the $Sp(2n), Sp(2n)/U(n)$ models respectively. 
Eq.~\ref{rgeqn}\ describes the evolution of
the physical coupling (and hence the spin conductance) with length
scale $L$, which could be either the system size or an inelastic
thermal cut-off length at finite temperature.  Note that in
two-dimensions ($\epsilon=0$), $g$ grows logarithmically with $L$,
giving an additive logarithmic reduction of the conductance and
signaling a cross-over to localized behavior at long distances.  
Notice that to this order (``one-loop'')  the leading
logarithmic correction is not completely suppressed by an orbital field. This
result is in sharp contrast to conventional weak-localization, but 
is in agreement with similar observations made in the context of the random matrix theory
of systems with these symmetries\cite{BB,AZ}. 
Complete suppression of the logarithmic correction occurs only with the
introduction of Zeeman coupling and subsequent crossover to the
unitary NL$\sigma$M.

\begin{figure}
\epsfxsize=2.0in
\centerline{\epsffile{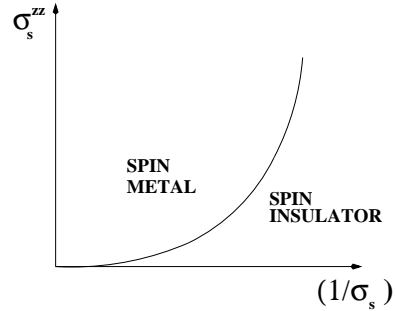}}
\vspace{0.05in}
\caption{Schematic phase diagram of the layered dirty $d_{x^2-y^2}$
superconductor. 
%$\sigma_s^{zz}$ and $\sigma_s$ are the bare spin
%conductivities perpendicular to the layers and of a single layer
%respectively.  Near the origin, the phase boundary obeys
%$\sigma_s^{zz} \sim e^{-4\pi^2
%\sigma_s}$
}
\vspace{0.04in}  
\label{smit}
\end{figure}

These perturbative results strongly suggest that in two spatial
dimensions with weak magnetic fields, the quasiparticles are
ultimately always localized.  A crude estimate for the localization
length may be obtained from the one loop perturbation theory to be
$\xi \sim \ell_e e^{\frac{4\pi}{g_0}}$ where $\ell_e$ is the mean free path
and $g_0$ the bare coupling constant. Using the estimate
$\frac{v_F}{v_{\scriptscriptstyle\Delta}} \sim 7$ in 
high-$T_c$, we get $\xi \approx 1000\ell_e$.
We note in passing that since $\Pi_2({Sp(2n) \over U(n)}) =Z$, a
non-trivial topological term is allowed for non-zero
orbital coupling; this suggests the possibility of isolated
extended quasiparticle states for strong magnetic fields\cite{long}. 

Inclusion of coupling between two-dimensional layers (with spacing
$d$) drives the system three dimensional, making possible an extended
phase where the spins diffuse at the longest length scales.  Based on
the quasi-2d NL$\sigma$M\cite{long}, the boundary between 3d
spin-metal and spin-localized phases occurs when the bare $z$-axis
spin conductivity $\sigma_s^{zz} \sim d\ell_e^{-2}e^{-8\pi/g}$ (see
Fig.~\ref{smit}).  Given this steep curvature of the phase boundary
near the origin, even a modest interlayer coupling can drive the
system into the spin metal phase.  In zero field, or 
neglecting Zeeman coupling, the
transition between the spin metal and the spin insulator is in a new
universality class.

Quasiparticle interactions, which we have ignored so far, can be
shown\cite{long}\ to lead to the usual Altshuler-Aronov singularities
for the tunneling density of states for the diffusive spin metal.
Interaction effects are expected to be more crucial in the spin
insulator, and ultimately should produce a low density (considerably
less than, e.g. the hole doping) of local magnetic moments which may
then at low temperature freeze into a spin glass or stay paramagnetic
in a random--singlet phase with a diverging spin susceptibility.

An important application of the theory outlined here is to the quantum
disordered $d_{x^2-y^2}$ superconductor -- a novel zero temperature
phase that has been proposed\cite{BFN}\ very recently to exist between
the antiferromagnetic and superconducting regions of the high-$T_c$
phase diagram.  The low temperature spin and thermal transport
properties of the nodal liquid are identical to that of the
superconductor, and with (weak) disorder, all the results mentioned
above obtain.  One quantitative point is worth mentioning: as one
moves from the superconductor towards the antiferromagnet through the
nodal liquid, the ratio $v_F/v_\Delta$ decreases monotonically, thereby
decreasing the bare spin conductance. Thus localization effects are
expected to become more important on going to the nodal liquid
region. It is interesting that it is precisely in this region that
experiments find a spin glass phase at low temperature.
  
We conclude with a few brief suggestions for experiments,
leaving a detailed discussion to Ref.~\onlinecite{long}.  Spin
transport can be probed by NMR techniques.  
It should also be possible to observe the localization physics in
thermal transport. Ignoring the weak interaction effects, we predict
that the thermal conductivity $\kappa$ is related to the {\em spin}
conductivity by the Wiedemann-Franz law:
\begin{equation}
\label{WF}
    \kappa/T\sigma_s = 4\pi^2/3.
\end{equation} 
Physically, this follows from the equality of spin and thermal
diffusion constants, the Einstein relation, and the relation between
specific heat and density of states. (The Lorenz number differs
by a factor of four from the usual one as the charge $e$ in the usual formula is
replaced by spin $\frac{1}{2}$ in our case.) Within the self-consistent theory
this gives $\kappa = 2k_B^2T(v_F^2 + v_{\Delta}^2)/3v_Fv_{\Delta}\hbar$. In contrast,
the microwave conductivity does not satisfy 
an Einstein relation and is, in general, not related to the thermal
conductivity by the Weidemann-Franz law (as can be explicitly seen in the
self-consistent theory, unless in the limit 
$v_F \gg v_{\scriptscriptstyle{\Delta}}$\cite{Graf}). 

Finally, we expect that {\sl localization} effects should be most
pronounced when the nodal anisotropy
$v_F/v_{\scriptscriptstyle\Delta}$ is minimized, as is expected to
occur on reducing the hole concentration within the nodal liquid
phase.  If signatures of localization can be observed, it may be
useful to perturb the system with a Zeeman (i.e. in-plane) field.  A
large enough Zeeman coupling is theoretically expected to open the
d-wave nodes into Fermi pockets\cite{Kun}, dramatically increasing the
density of states and the bare conductance, hence potentially probing
some of the localization transitions discussed here.

We thank  Ilya Gruzberg, Doug Scalapino, and Kun Yang for useful discussions,
and Martin Zirnbauer for useful comments on the manuscript. 
This research was supported by NSF Grants DMR-97-04005,
DMR95-28578
and PHY94-07194.
We have recently come across papers by Bundschuh
et. al.\cite{Carlos} discussing quasiparticle localization in
vortex cores of superconductors using a supersymmetric
formalism. In the region of overlap, their results agree with those in this paper.


\vskip -0.2in

\begin{references}
\bibitem{lucent} Present address: Rm. 1D-368,
Lucent Technologies, Bell Labs, 700 Mountain Ave., Murray Hill, NJ
07974

\bibitem{ucla} Present address:Physics Department, University of California, 
Los Angeles, CA
90095--1547

 
 
\bibitem{Kirtley}D.A. Wollman et. al., Phys. Rev. Lett. {\bf 74}, 797 (1995);
C.C. Tsuei et. al., Phys. Rev. Lett.{\bf 73},593 (1994)

\bibitem{Lee} P.A. Lee, Phys. Rev. Lett., {\bf 71}, 1887 (1993)

\bibitem{Gorkov} L.P. Gorkov and P.A. Kalugin, JETP Lett. {\bf 41},253 (1985); 
S.Schmitt-Rink {\it et. al.}, Phys. Rev. Lett. {\bf 57},
2575 (1986)

\bibitem{HPS} P.J. Hirschfeld {\it et. al.}, Phys. Rev. {\bf B50},
10250 (1994)

\bibitem{Graf} M. Graf et. al., Phys. Rev. {\bf B53}, 15147 (1996)

\bibitem{AZ} A. Altland and M.R. Zirnbauer, Phys. Rev. {\bf B 55}, 1142 (1997);
M.R. Zirnbauer, J. Math. Phys.,{\bf 37}, 4986 (1996)

\bibitem{BFN}L. Balents, M.P.A. Fisher and C. Nayak, cond-mat/9803086

\bibitem{Tsvelik} See, e.g. H. E. Castillo {\it et. al.}, Phys. Rev. B
{\bf 56}, 10668 (1997); J. S. Caux {\it et. al.},
Phys. Rev. Lett. {\bf 80}, 1276 (1998); and references therein.



\bibitem{long} T. Senthil, L. Balents, M. P. A. Fisher, and 
C. Nayak, in preparation.


 
\bibitem{Polyakov} A.M. Polyakov, {\sl Gauge fields and strings}
(Harwood, London); F. Wegner, Nucl. Phys. {\bf B316}, 663 (1989)

\bibitem{BB} P.W. Brouwer and C.W.J. Beenakker, Phys. Rev. {\bf B 52}, 3868 (1995)

\bibitem{Kun} Kun Yang and S. Sondhi, Phys. Rev. {\bf B57}, 8566 (1998)

\bibitem{Carlos} R. Bundschuh et. al., cond-mat/9806172; cond-mat/9808297 
\end{references}
\end{document}